\documentclass[aps,prb,floatfix,twocolumn,amsmath,amssymb,showpacs,10pt]{revtex4-1}

\usepackage{amsmath}  
\usepackage{amsfonts} 
\usepackage{graphicx} 
\usepackage{pst-all,pstricks-add}
\usepackage{dcolumn}

\newcommand{\myvector}[1]{\boldsymbol{\mathrm{#1}}}

\newcommand{\unit}[1]{\, \textrm{#1}}

\begin{document}


\title{Connecting optical intensities and electric fields using a triple interferometer}

\author{David Collins}
\email{dacollin@coloradomesa.edu} 
\affiliation{Department of Physical and Environmental Sciences, Colorado Mesa University, 1100 North Avenue, Grand Junction, CO 81501}

\author{Justin Endicott}
\altaffiliation[Current Affiliation: ]{Blue Line Engineering, 525 E. Colorado Avenue, Colorado Springs, CO 80903}
\affiliation{Department of Physical and Environmental Sciences, Colorado Mesa University, 1100 North Avenue, Grand Junction, CO 81501}

\date{\today}

\begin{abstract}
We consider the issue of validating the relationship between electric fields and optical intensity as proposed by the classical theory of electromagnetism. We describe an interference scenario in which this can be checked using only intensity measurements and without any other information regarding the details of the arrangement of the associated fields. We implement this experimentally using a triple Michelson interferometer and the results strongly suggest that the method validates the classical relationship between optical intensity and the associated classical field. 
\end{abstract}

\maketitle 

\section{Introduction} 
\label{sec:intro}

A central tenet of classical optics is that light can be described via an associated electric field. The behavior of the electric field is determined by the theory of electromagnetism, which eventually relates the intensity of a light source to the associated electric field~\cite{brooker02, hecht02,bennett08, kenyon08}. The process of validating the electric field description of light is complicated by the apparent difficulty of measuring the associated electric fields directly. The theory is usually checked indirectly via inferences based on intensity measurements. 

One such indirect inference involves light produced from two or more sources. The associated electric fields interfere, producing a superposition field which determines the observed intensity. The resulting interference phenomena, especially those produced by single or multiple slits or interferometers, and their relationships with electric fields  are familiar to most undergraduate physics students~\cite{brooker02, hecht02,bennett08, kenyon08}. In general, a detailed analysis of such interference phenomena is not done purely in terms of intensities but also involves phase relationships between the individual sources. This typically depends very delicately on the configuration of the sources and when many of these are present a precise comparison between experimental results and theoretical predictions can be difficult. 

However, recently there has emerged a type of multipath interference scenario which only uses intensity information to explore and validate an analogous quantum theory model of light~\cite{sorkin94}. This has been checked with experiments involving superposition of light from more than two sources~\cite{sinha10,sollner12,kauten17}, produced by special multiple slit and mask arrangements. These experiments appeared to validate the underlying theory, when applied in a simplified fashion. However, a more detailed analysis revealed a small discrepancy~\cite{deraedt12,sawant14,quach17,barnea18,skagerstam18} and these deviations have been checked experimentally~\cite{rengaraj18}. Similar theoretical and experimental investigations into analogs of multipath interference have been conducted in situations that do not involve optical slits~\cite{jin17,cotter19,lee20,pleinert20}.

In this article we describe an adaptation of such multipath experiments that checks the relationship between classical electric fields and optical intensity and a method for assessing interference when multiple sources are present. Rather than a multiple slit arrangement~\cite{sinha10,sollner12}, the experiment uses a triple interferometer. This has the advantage of being much easier to manage than the comparable multiple slit experiments and is within the abilities of undergraduate students. It would also introduce undergraduate students to situations in which inferences are made via correlating results (in this case intensities) from various experimental settings and observing how this technique can illuminate the underlying physics. Finally it also appears to evade the immediate critique~\cite{deraedt12} of previous multipath experiments.    

This article is organized as follows. Section~\ref{sec:theory} describes the theoretical background and how interference is quantified. Section~\ref{sec:expsetup} describes the experimental setup and section~\ref{sec:results} describes the results of the experiment. 


\section{Intensity and Interference Terms}
\label{sec:theory}

In optics, the intensity (or irradiance) of light is defined to be the time averaged rate at which energy flows across a surface per unit area of that surface~\cite{brooker02, hecht02,bennett08, kenyon08}. In classical physical optics light is described as an electromagnetic wave and the rate of energy propagation per unit area is determined by the time average of the Poynting vector associated with the electromagnetic field. The simplest case to assess is that of a monochromatic electromagnetic field whose electric field at location $\myvector{r}$ and time $t$ is $\myvector{E} = \myvector{E}_0 \cos{\left( \myvector{k}\cdot \myvector{r} - \omega t + \phi \right )}$; here $\myvector{k}$ is the wavenumber vector associated with the direction of propagation and wavelength of the wave, $\omega$ is the angular frequency of the wave, $\phi$ is the phase of the wave, and $\myvector{E}_0$ is a vector that is independent of location and time. It emerges~\cite{hecht02,bennett08} that, in free space, the intensity of this wave is $I = \epsilon_0 c E_0^2/2$ where $\epsilon_0$ is the permittivity of free space. In the complex formalism, this electromagnetic wave is described via $\myvector{E} = \myvector{E}_0 e^{i(\myvector{k}\cdot \myvector{r} - \omega t + \phi)}$ and then the intensity is 
\begin{equation}
I = \frac{\epsilon_0 c}{2} \myvector{E}\cdot \myvector{E}^*.
 \label{eq:classicalintensity}
\end{equation} 
Crucially the intensity is proportional to $\myvector{E}\cdot \myvector{E}^*$. 

Now suppose that two sources, A and B, each produce light with the same wavenumber and the superposition is incident on a detector. In classical electromagnetism the electric fields produced by various sources are superposed linearly to form the field that will be detected by any detector. The electric field arriving at the detector is $\myvector{E} = \myvector{E}_A + \myvector{E}_B$ where $\myvector{E}_A$ is the electric field produced by source A and $\myvector{E}_B$ is that produced by source B. The resulting intensity is
\begin{equation}
 I = I_A + I_B + \frac{\epsilon_0 c}{2} \left( \myvector{E}_A\cdot \myvector{E}_B^* + \myvector{E}_B\cdot \myvector{E}_A^* \right),
 \label{eq:twosourceintensity}
\end{equation} 
where $I_A = \epsilon_0 c \myvector{E}_A\cdot \myvector{E}_A^*/2$ is the intensity if only source A were present. $I_B$ is defined similarly. If $\myvector{E} = \myvector{E}_{0A} e^{i(\myvector{k}\cdot \myvector{r} - \omega t + \phi_A)}$ where $\myvector{E}_{0A}$ and $\phi_A$ are constant and a similar expression applies to source B, then straightforward analysis gives
\begin{equation}
 I_{AB} = I_A + I_B + 2 \sqrt{I_A I_B} \cos{\left( \Delta \phi \right)}
 \label{eq:twosourceintensitytwo}
\end{equation}
where $\Delta \phi := \phi_A - \phi_B$ is the phase shift between the sources. The resulting interference between the sources is a key prediction that results from the underlying electric field description and is at odds with a simplistic description which would assume that the intensity of the combination is the sum of the two intensities. However, checking a prediction of Eq.~\eqref{eq:twosourceintensitytwo} requires knowing the phase shift between the sources. This can depend in a complicated way on the configuration of the sources. 

This complication can be avoided by considering multipath interference experiments~\cite{sinha10,sollner12} which typically combine light produced by multiple sources at a detector. We initially develop the associated theory purely in terms of intensities, ignoring whatever underlying theory may describe these intensities. 

Suppose that there are three sources, labeled A, B and C. Each source can be turned on and off independently and at will and whenever any source is turned on it produces light with a set constant intensity; thus whenever source A is turned on it produces light with the same intensity as whenever it had been turned on previously. The detector only measures the overall intensity of the light that arrives at it resulting from all the sources. We use the following notation to describe the possible intensities recorded by the detector when various sources are are on or off.  Let $I_A$ be the intensity recorded by the detector when source A is on and sources B and C are off. Let $I_B$ be the intensity recorded by the detector when source B is on and sources A and C are off. Define $I_C$ similarly. Then let $I_{AB}$ be the intensity recorded by the detector when sources A and B are on and source C is off. Define $I_{AC}$ and $I_{BC}$ in a corresponding fashion. Let $I_{ABC}$ be the intensity recorded by the detector if all three sources are on and $I_0$ be the intensity if none are turned on. The central questions ask how $I_{ABC}$ is related to $I_{AB}, I_{BC}, I_{AC}, I_{A}, I_{B}, I_{C}$ and $I_{0}$ or how $I_{AB}$ is related to $I_{A}, I_{B}$ and $I_0$.

Such issues have been addressed in the context of various probabilistic descriptions and measures within quantum theory~\cite{sorkin94} and can be adapted to classical optics. Without knowing any details of the underlying theory that describes the intensity we could entertain various possibilities. For example, if the theory were such that the intensities superimposed linearly, then $I_{AB} = I_A + I_B$. This motivates the definition of a second order interference term~\cite{sorkin94},
\begin{equation}
 \Delta_2(A,B) := I_{AB} - I_A - I_B.
\end{equation}
The quantities on the right can be measured experimentally regardless of the underlying theory that describes the values of that on the left. Various theoretical models could then predict $\Delta_2(A,B)$ and checked against value computed via measurements. For example, if the intensities superimposed linearly then $\Delta_2(A,B)=0.$

According to classical electromagnetism, Eq.~\eqref{eq:twosourceintensitytwo} predicts that 
\begin{equation}
 \Delta_2(A,B) = 2 \sqrt{I_A I_B} \cos{\left( \Delta \phi \right)}
\end{equation}
and the sources could always be arranged with a phase shift such that $\Delta_2(A,B) \neq 0$. At this point given a choice between a theory in which the intensities superimpose linearly and one in which fields superimpose linearly, measuring the intensities and computing $\Delta_2(A,B)$ would allow us to decide which of these two possibilities would be correct. 

However, if we cannot measure the electric fields directly or if it is difficult to ascertain or control the  phase shift, then we cannot use intensity measurements to easily check the predictions of classical electromagnetic theory. We therefore seek a comparable quantity which will allow us to check the predictions of classical electromagnetism  only using intensity measurements.

If all three sources are turned on then, $\myvector{E} = \myvector{E}_A + \myvector{E}_B + \myvector{E}_C$ and
\begin{eqnarray}
 I_{ABC} & = & \frac{\epsilon_0 c}{2} 
               \left( \myvector{E}_A + \myvector{E}_B + \myvector{E}_C
               \right)
							 \cdot 
               \left( \myvector{E}_A^* + \myvector{E}_B^* + \myvector{E}_C^*
               \right)
							 \nonumber \\
				& = & I_A + I_B + I_C 
				     + \frac{\epsilon_0 c}{2} 
               \left( \myvector{E}_A \myvector{E}_B^* + \myvector{E}_A^* \myvector{E}_B
							 \right) 
				      \nonumber \\
				&   & + \frac{\epsilon_0 c}{2} 
               \left( \myvector{E}_A \myvector{E}_C^* + \myvector{E}_A^* \myvector{E}_C
							        + \myvector{E}_B \myvector{E}_C^* + \myvector{E}_B^* \myvector{E}_C
							 \right)
							 \nonumber \\
				& = & I_A + I_B + I_C + I_{AB} - I_A - I_B 
				      \nonumber \\
				&   & + I_{AC} - I_A - I_C + I_{BC} - I_B - I_C
							 \nonumber \\
				& = & I_{AB} + I_{AC} + I_{BC} - I_A - I_B - I_C. 
 \label{eq:thirdorderitderiv}
\end{eqnarray}
It follows that, regardless of the intensities of the individual sources or the phase relationship between the associated electromagnetic waves it is always true that
\begin{equation}
 I_{ABC} - I_{AB} - I_{AC} - I_{BC} + I_A + I_B + I_C = 0.
\end{equation}
We then define a third order interference term~\cite{sorkin94}, also called the Sorkin parameter,
\begin{equation}
 \Delta_{3}(A,B,C) := I_{ABC} - I_{AB} - I_{AC} - I_{BC} + I_A + I_B + I_C.
 \label{eq:thirdorderit}
\end{equation}
Then if, as classical electromagnetism predicts, the intensity is determined via Eq.~\eqref{eq:classicalintensity} and the electric fields superimpose linearly, then $\Delta_{3}(A,B,C)=0$ but, in general, $\Delta_{2}(A,B) \neq 0$, $\Delta_{2}(B,C) \neq 0$ and $\Delta_{2}(A,C) \neq 0$. 

We briefly consider the possibility that all second order interference terms are zero. A second order interference term is only zero if and only if the phase shift between the two sources is an odd multiple of $\pi/2$. However, if the phase shift between A and B is an odd multiple of $\pi/2$ and the same is true for that between A and C, then the phase shift between B and C is an even multiple of $\pi/2$. Thus it is impossible that all three second order interference terms are zero. 

Thus if the predictions of classical electromagnetism are correct then $\Delta_{3}(A,B,C)=0$ and at least one second order interference term is non-zero. Note that this method for checking the underlying electric field description is insensitive to the intensities of the individual sources and the phase relationship between them. 

This third order interference term can be expressed in terms of second order interference terms such as
\begin{equation}
 \Delta_{2}(AB,C)  = I_{ABC} - I_{AB} - I_C
\end{equation}
and it is easily seen that, for example, 
\begin{equation}
 \Delta_{3}(A,B,C)  = \Delta_{2}(AB,C) - \Delta_{2}(A,C) - \Delta_{2}(B,C).
\end{equation}
It immediately follows that any theory for which the second order interference term is always zero implies that the third order interference term would also be zero; an example would be a theory in which the intensities superimpose linearly. The converse is clearly not true; one counterexample is classical electromagnetism and optics.

This entire framework has been extended~\cite{sorkin94} to arbitrarily high order interference terms and has the feature that any theory in which the interference term at a given order is zero automatically implies that higher order interference terms are zero. Finding the boundary between the interference terms which are zero and those which are not then delimits the possible theory. In the case of classical electromagnetism and optics the boundary is between the second and third order. The experiment aims to check this.


\section{Experimental Set-Up}
\label{sec:expsetup}

Previous experiments which have investigated interference in optics have used multiple slits to act as the required sources~\cite{sinha10}. These used a succession of single photons, followed by photon counting to check intensity predictions given via the Born rule. However, it emerged that a detailed theoretical analysis of the intensities produced by various slit arrangements yields a small non-zero third order interference term~\cite{deraedt12} and thus the fields produced in this way do not superimpose exactly as the model that yields $\Delta_{3}(A,B,C)=0$ predicts. 

Additionally these experiments require delicate manipulation of a closely spaced multiple slit arrangement and the masks which open or close various slits as well as intricacies associated with generating and counting single photons. The experiment that we describe avoids these technical issues but still illustrates how the hierarchy of interference terms can decide between various theories. 

\begin{figure}[h!]%
 \includegraphics[scale=1]{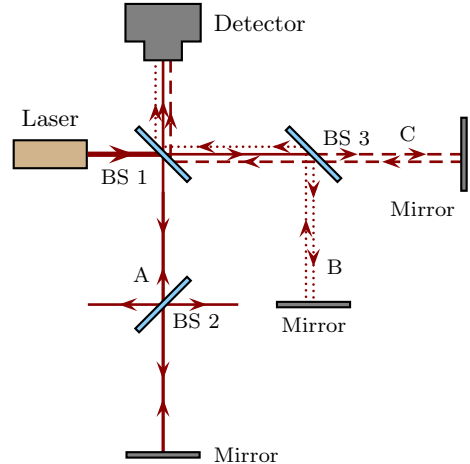}	
\caption{Triple Michelson interferometer. Light from a source is incident on beam splitter (BS 1), which forms the parent Michelson interferometer. The resulting transmitted and reflected beams are incident on two other beam splitters (BS~2 and BS~3), which initiate the offspring interferometers. The reflected and transmitted beams from BS~3 are redirected via mirrors, eventually reaching the detector. Prior to the detector they form source B (dotted) and source C (dashed). The beam transmitted through BS~2 is also redirected via a mirror, eventually forming source A (solid) prior to the detector. The horizontal beams from BS~2 are discarded.  
				\label{fig:tripleinterferometer}%
				}
\end{figure}

Our experiment uses a triple Michelson interferometer to produce three sources. This interferometer consists of a parent interferometer and two offspring interferometers configured as illustrated in Fig.~\ref{fig:tripleinterferometer}.

This arrangement can effectively produce beams from three sources incident on the detector. Sources can be turned on and off by blocking the relevant arms within the interferometer. Note that BS~2 in Fig.~\ref{fig:tripleinterferometer} is not strictly necessary for the production of source A. However, it does allow for a situation where the intensity of all three sources is approximately equal and where the appearance of the interference pattern at the detector is roughly equally sensitive to an adjustment in either offspring interferometer. 

The beam generation was done using a Melles Griot 25-LHP-111-249 $1.0\unit{mW}$ HeNe laser as source. The three beam splitters were each Thorlabs BSW10 plate beam splitters, which are nominally 50:50. The mirrors are Thorlabs BB1-E01 broadband mirrors. One of the beam splitters and all three mirrors were mounted in Thorlabs KM100 kinematic mirror mounts to allow for beam alignment. Beam paths in the arms of the interferometer were block by Thorlabs BB1-E01 broadband mirrors arranged to redirect beams perpendicular to the interferometer plane (out of the page in Fig.~\ref{fig:tripleinterferometer}). These were mounted to Thorlabs TRF90 flip mounts to allow for beams to be turned on and off easily. A lens was inserted between BS~1 and the detector to expand the interference pattern produced by the three beams at the detector. This had two purposes. The first was to render the pattern fringes visible to the naked eye and assist in the beam alignment. The second was to allow for detection at various locations along the fringe pattern and thereby permit various relative phase shifts between the effective sources.

\begin{figure}[h!]%
 \includegraphics[scale=0.50]{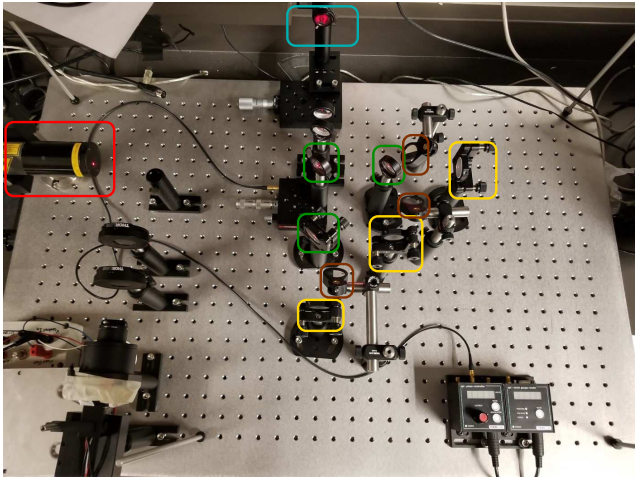}	
\caption{Experimental set up. The red box indicates the laser source, the blue the detector, the green the beamsplitters, the yellow the fixed mirrors and the brown the flip mirrors.   
				\label{fig:apparatus}%
				}
\end{figure}

The detector was a Thorlabs DET36A biased photodector (PD), whose current output is linearly proportional to the input optical power.  A Thorlabs FB630-10 bandpass filter (central wavelength $630\pm2\unit{nm}$ and FWHM $10\pm 2\unit{nm}$) was mounted to the detector in order to reduce the residual stray ambient light incident on the detector. This was preceded by a Thorlabs SM1D12 iris diaphragm, whose aperture was adjusted to its minimum diameter (nominally $1\unit{mm}$) so as to restrict collection of light from a small portion of the interference pattern. The detector was mounted onto a Thorlabs linear stage oriented to allow for motion perpendicular to the beam direction. This allowed us to explore various parts of the interference pattern. The entire arrangement was mounted to a Thorlabs PBH11106 breadboard. The setup is illustrated in Fig.~\ref{fig:apparatus}. 

The voltage output from the PD was acquired via a PASCO 550 Universal Interface with a PASCO voltage sensor. The PD voltage output was recorded as a function of time and displayed using PASCO Capstone software.

The experiment was conducted in a dark room but this did not eliminate all intrusion of light produced from sources other than the laser and the PD would provide a small non-zero voltage reading even when all three sources were blocked. The associated background intensity $I_0$ must be subtracted from every intensity that enters into Eq.~\eqref{eq:thirdorderit}. The result is a modified version of the third order interference term, 
\begin{equation}
 \Delta_3(A,B,C) = I_{ABC} - I_{AB} - I_{AC} - I_{BC} + I_A + I_B + I_C - I_0.
 \label{eq:expdeltathree}
\end{equation}
We then aim to verify whether 
\begin{equation}
 \Delta_3(A,B,C) = 0
\end{equation}
provided that the intensities are those measured by the PD. Additionally note that since the intensity of the light incident on the PD is proportional to the PD output, converted into a voltage, we can replace the intensities in Eq.~\eqref{eq:expdeltathree} by the associated voltages and will do so for the remainder of this article. 

A single ``setting'' of the experiment consisted of the following sequence. 

\begin{enumerate}
 \item Position the detector at a fixed location along the interference pattern.
 \item Allow all three beams to be incident and record the PD output. This gives $I_{ABC}$.
 \item Block path C and record the PD output. This gives $I_{AB}.$
 \item Open path C and block B and record the PD output. This gives $I_{AC}.$
 \item Open path B and block A and record the PD output. This gives $I_{BC}.$
 \item Block B and C and record the PD output. This gives $I_{A}.$
 \item Block A and C and record the PD output. This gives $I_{B}.$
 \item Block A and B and record the PD output. This gives $I_{C}.$
 \item Block all three paths and record the PD output. This gives $I_{0}.$
\end{enumerate}

Five runs were done at each setting. A total of eleven different settings were used, corresponding to eleven different detector positions along the interference pattern. This effectively samples eleven different phase relationships between the three beams.


\section{Data and Results}
\label{sec:results}

\begin{figure}[h!]%
 \includegraphics[scale=0.70]{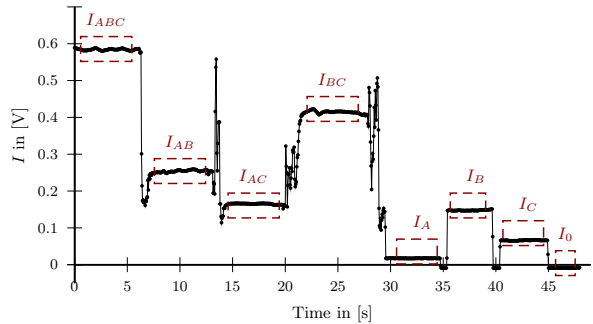}	
\caption{PD output voltage versus time for one particular single run. The boxes indicate the moments during which various paths were opened or closed. During these periods the intensity was more or less constant. The vertical lines and spikes appear while the flip mirrors are being adjusted to toggle between sources.   
				\label{fig:clean}%
				}
\end{figure}

\begin{figure}[h!]%
 \includegraphics[scale=0.70]{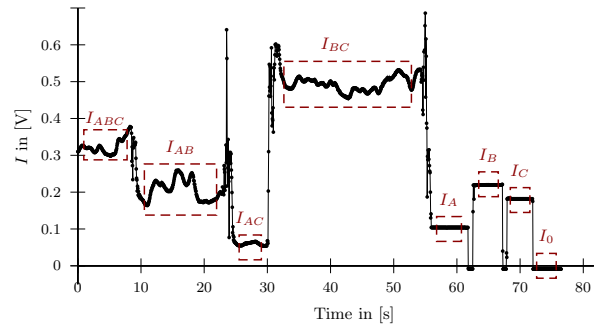}	
\caption{PD output voltage versus time for one particular single run that yielded messier data. Symbols have the same meaning as in Fig.~\ref{fig:clean}.   
				\label{fig:messy}%
				}
\end{figure}

Capstone recorded the voltage produced by the PD as a function of time continuously during each run of the experiment.  Typical examples are illustrated in Fig.~\ref{fig:clean}, representative of a cleaner data set, and Fig.~\ref{fig:messy}, representative of a noisier data set. In each run the steep vertical transitions and spikes indicate the moments during which the flip mirrors are moved so as to alter the beam combination incident on the detector. These then delineate intervals during which the intensity is produced by particular combinations of beams. Each figure shows eight such intervals, each typically lasting for five to ten seconds. During each interval, the intensity should be constant although the degree to which this occurred varied. A representative intensity for each interval was determined via the mean and standard deviation of all voltages spanning the period between transitions but excluding buffer periods of approximately equal duration (one or two seconds) before and after the transitions. 

The collection of eight such data points for one run were substituted into Eq.~\eqref{eq:expdeltathree} to determine $\Delta_3(A,B,C), \Delta_2(A,B), \Delta_2(A,C)$ and $\Delta_2(B,C)$. For each setting, the data was used to determine a weighted average for all interference terms and these are displayed in Table~\ref{tab:data}.

\onecolumngrid

\begin{table}[h!]
\centering
\begin{ruledtabular}
\begin{tabular}{ccccc}
 Setting & $\Delta_3(A,B,C)$ & $\Delta_2(A,B)$ & $\Delta_2(A,C)$ & $\Delta_2(B,C)$\\
 \hline
 1 & $2.4 \pm 3.3 \unit{mV}$ & $74.5 \pm 1.0 \unit{mV}$ & $68.3 \pm 1.1 \unit{mV}$ & $179.7 \pm 0.8 \unit{mV}$ \\
 2 & $-16 \pm 9 \unit{mV}$ & $98.0 \pm 1.8 \unit{mV}$ & $73.7 \pm 2.5 \unit{mV}$ & $220.8 \pm 1.9 \unit{mV}$ \\
 3 & $-2 \pm 10 \unit{mV}$ & $117.7 \pm 2.3 \unit{mV}$ & $104.1 \pm 1.9 \unit{mV}$ & $225 \pm 4 \unit{mV}$ \\
 4 & $-2.6 \pm 2.3 \unit{mV}$ & $-59 \pm 10 \unit{mV}$ & $-194.2 \pm 0.8 \unit{mV}$ & $76 \pm 11 \unit{mV}$ \\
 5 & $10 \pm 21 \unit{mV}$ & $-131 \pm 8 \unit{mV}$ & $-227 \pm 4 \unit{mV}$ & $64 \pm 12 \unit{mV}$ \\
 6 & $-3 \pm 5 \unit{mV}$ & $-212.8 \pm 1.7 \unit{mV}$ & $-290.4 \pm 1.2 \unit{mV}$ & $386.0 \pm 2.0 \unit{mV}$ \\
 7 & $23 \pm 10 \unit{mV}$ & $-216 \pm 3 \unit{mV}$ & $-335.9 \pm 1.9 \unit{mV}$ & $278.0 \pm 3.0 \unit{mV}$ \\
 8 & $26 \pm 18 \unit{mV}$ & $136 \pm 7 \unit{mV}$ & $46.3 \pm 2.0 \unit{mV}$ & $-409.2 \pm 2.4 \unit{mV}$ \\
 9 & $-12 \pm 17 \unit{mV}$ & $-265.4 \pm 0.9 \unit{mV}$ & $442 \pm 6 \unit{mV}$ & $-387 \pm 6 \unit{mV}$ \\
 10 & $-0.0 \pm 2.0 \unit{mV}$ & $-158 \pm 8 \unit{mV}$ & $438 \pm 8 \unit{mV}$ & $-299 \pm 3 \unit{mV}$ \\
 10 & $-7 \pm 18\unit{mV}$ & $23 \pm 4 \unit{mV}$ & $488 \pm 5 \unit{mV}$ & $190 \pm 7 \unit{mV}$ \\
\end{tabular}
\end{ruledtabular}
\caption{Data for interference terms averaged for each setting. \label{tab:data}}
\end{table}

\twocolumngrid

\begin{figure}[h!]%
 \includegraphics[scale=0.60]{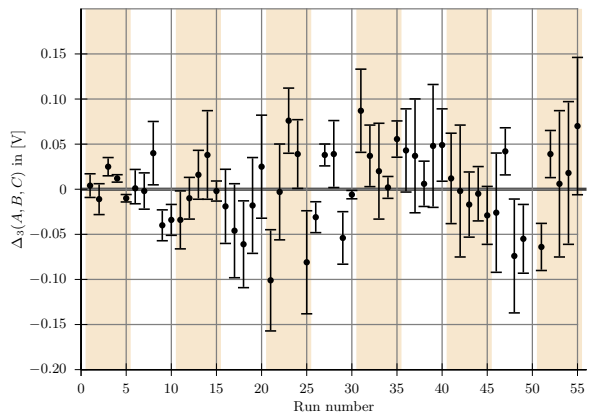}	
\caption{Data for $\Delta_3(A,B,C)$. The intensities are represented by the voltage readings produced by the PD in V. The white and color bands delineate runs with the same setting along the interference pattern.  
				\label{fig:deltathreedata}%
				}
\end{figure}

\begin{figure}[h!]%
 \includegraphics[scale=0.60]{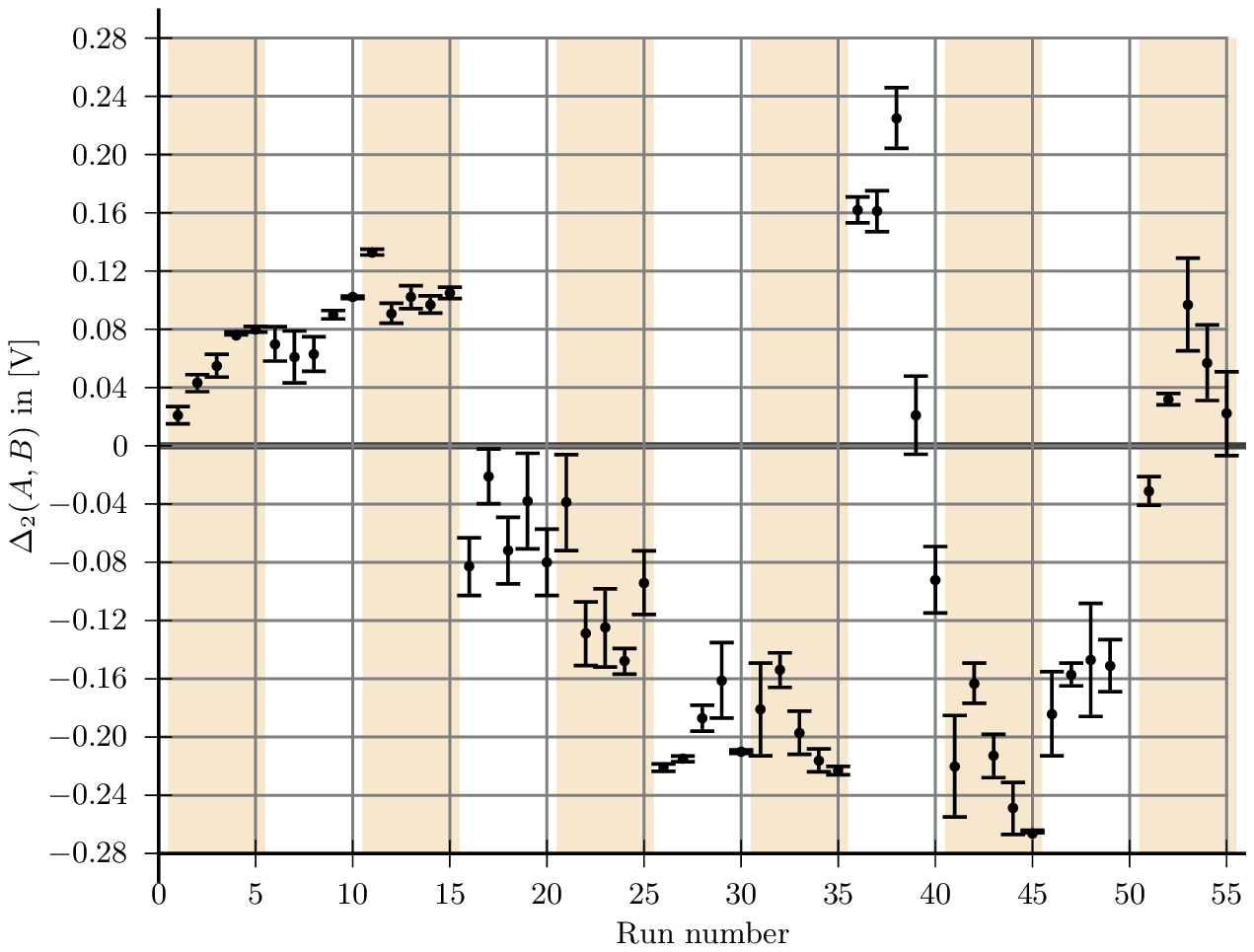}	
\caption{Data for $\Delta_2(A,B)$ with a set-up similar to Fig.~\ref{fig:deltathreedata}.   
				\label{fig:deltatwoabdata}%
				}
\end{figure}

\begin{figure}[h!]%
 \includegraphics[scale=0.60]{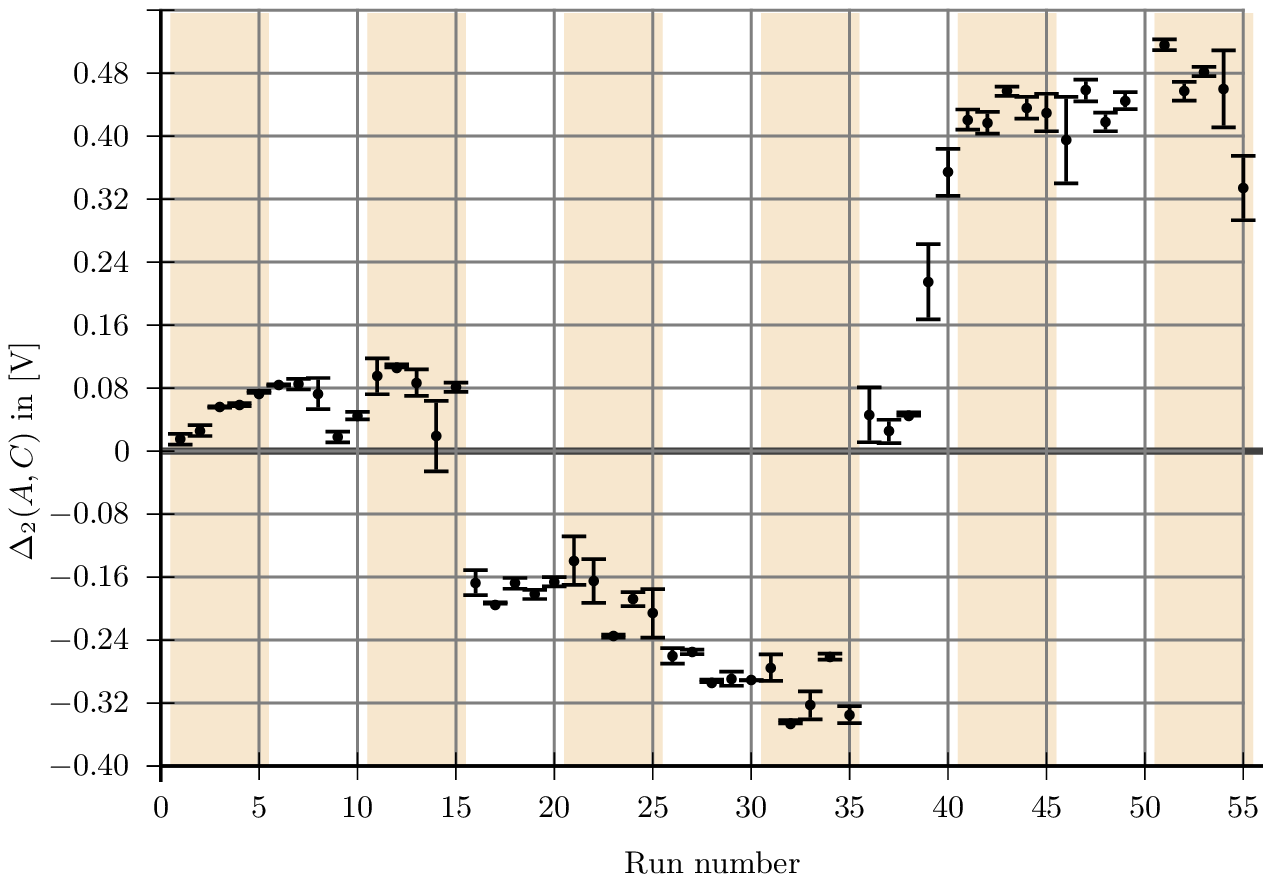}	
\caption{Data for $\Delta_2(A,C)$  with a set-up similar to Fig.~\ref{fig:deltathreedata}.  
				\label{fig:deltatwoacdata}%
				}
\end{figure}

\begin{figure}%
 \includegraphics[scale=0.60]{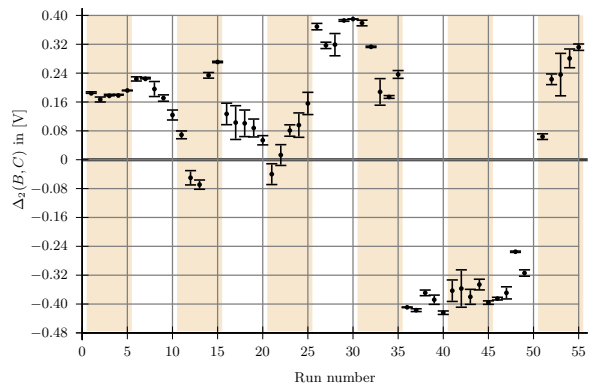}	
\caption{Data for $\Delta_2(B,C)$  with a set-up similar to Fig.~\ref{fig:deltathreedata}.    
				\label{fig:deltatwobcdata}%
				}
\end{figure}

The data for $54$ runs are displayed in Figs.~\ref{fig:deltathreedata}--\ref{fig:deltatwobcdata} (the data from one run had been inadvertently erased).

Of the eleven settings, the third order interference term is within one standard deviation of $0\unit{mV}$ for seven. For three of the settings it is within one standard deviation and for one setting it is within two. This should be compared against the second order terms for each setting. According to classical electromagnetic theory and Eq.~\eqref{eq:twosourceintensitytwo} the second order term could be $0\unit{mV}$ whenever $\Delta \phi$ is an odd half multiple of $\pi/2.$ The experiment did not attempt to control the relative phases and it would have been possible for at least one of the second order terms to be $0\unit{mV}$. However, all three second order terms cannot be $0\unit{mV}$. In general our data shows that none of the second order terms is within five standard deviations of $0\unit{mV}$. Furthermore, considering each run there is always at least one second order interference term which is beyond $50$ standard deviations from $0\unit{mV}$ (the lowest such maximum occurs for setting 3). This and the fact that the majority of the settings yielded a third order term within a standard deviation of $0\unit{mV}$ strongly suggest that this experiment validates the predictions of classical electromagnetism with regard to the relationship between optical intensities and electromagnetic fields. 

The primary source of error in these experiments is most likely the ability to produce stable interference between the beams that are incident on the detector. Figures~\ref{fig:clean} and~\ref{fig:messy} show that when only one source is turned on (giving $I_A, I_B$ or $I_C$) the signal produced by the PD is fairly stable. On the other hand when two or three sources are turned on the signal becomes less stable; Fig.~\ref{fig:messy} illustrates this. This probably is a result of the fact that either five or six optical components are involved in the production of the signal that arrives at the source. Fluctuations on the order of the wavelength of the light will clearly dramatically alter the resulting interference pattern. Such fluctuations can easily result from vibrations as the optical bench was not isolated or thermal drift, which would alter the index of refraction and therefore the relative phase shifts along the paths. In fact, we noticed that when aligning the optics the qualitative appearance of the interference pattern was extremely sensitive to adjustments to mirrors or beamsplitters. In preliminary attempts to gain data we also observed that the airflow provided by the room ventilation system created a noticeable drift in the visible interference pattern. This airflow was eliminated and that data was excluded from consideration but this illustrates the difficulty of producing stable interference patterns in this type of triple Michelson interferometer. 

Note that the analysis via interference terms does not require knowledge of the phase shifts between the three sources. These phase shifts depend on the precise alignment of the mirrors and beamsplitters and the position of the detector. We found that, during optical element alignment, the visual appearance of the interference pattern was very sensitive to adjustments and we doubt that we could have predicted the associated phase shifts. The relative strengths of the the electric fields produced by the three sources also depends on the reflectivity and transmittivity of the beamsplitters and mirrors. Again these details are irrelevant for the analysis in terms of the interference terms. All that is required are the various intensities at the detector for each setting.  

\section{Discussion and Conclusion}

We have presented an experiment to validate the description of optical intensity via electromagnetic fields associated with light. The experiment only relies on intensity measurements and does not require any information about the relative phases between multiple sources that produce the light that is subsequently detected. The experiment introduces measurable interference terms and relies on these to validate whether the usual theory is correct. The resulting data strongly suggests that the experiment has established the validity of this approach.     

A strength of this approach is that it does not rely on knowledge of the precise phase relationship between light sources that superimpose. This knowledge is essential with typical investigations of interference using multiple slits or even interferometers. The relative phases between sources are invariably very sensitive to adjustments and the elimination of this issue vastly simplifies the experiment. 

We believe that the experiment is easier to manage than comparable experiments involving multiple slit interference~\cite{sinha10,sollner12} and the particular critique of those experiments whose conclusion is that the third order interference term is non-zero does not immediately apply to our experiment~\cite{deraedt12}. Whether a comparable issue might arise for our type of experiment is an open question. 

This experiment could be extended in various ways. First, it could be done with true single photon sources and use photon counting rather than intensity measurements. In this way the rules connecting quantum states and probabilities could be checked; this was done in the multiple slit experiments~\cite{sinha10,sollner12}. This would entail the cost and management of single photon sources and photon counting devices and would also introduce statistical analyses associated with dark counts and detector efficiencies. However, for undergraduate students, it would offer the benefit of direct use of the foundations of quantum theory to predict the outcome of experiments.

Second, the layout of the experiment allows for introduction of additional optical elements into individual beam paths. This could be used to incorporate the effects of polarization of the light sources. For example, if the polarization state of one source could be rotated relative to the others then the relationships that the interference terms satisfy would change. These could be explored in an experiment where the data gathering is no more difficult than that which we have done. At the classical level, this would expose undergraduate students to the vector nature of the electric field and its relationship to intensity. At the quantum level, it would allow students to explore the quantum nature of the path possibilities alongside that of polarization. Such investigations would be very difficult to conduct with the multiple slit and mask arrangement of previous experiments~\cite{sinha10,sollner12}. 

This type of analysis allows for testing of candidate theories beyond those of simple addition of intensity or that resulting from classical electromagnetic fields. For example, perhaps a possible theory predicts that the intensity produced when three sources superimpose is a combination of the intensities for all twofold combinations and does not depend on the intensities when a single source is active. With energy conservation, such a theory might predict that $I_{ABC} = \left( I_{AB} + I_{BC} + I_{AC} \right)/2$. Although there is no reason to expect such a theory to be true, our experiment and data could easily check this. In such a way a myriad of theories that make predictions about intensities could be checked via the appropriate combinations of intensities. 

The forgoing is analogous to the type of thinking used in Bell inequality experiments~\cite{mermin81,aspect81,giustina15} that compare predictions of quantum physics to a broad class of plausible physical theories. Rather than try to investigate the properties of quantum states or competing alternatives directly, such experiments use probabilities and correlations, toggling between a variety of experimental settings and ultimately combining the resulting probabilities is a sensible way. Our experiment has the same flavor and we think would be very instructive way for undergraduate students to explore similar types of indirect inferences.

\end{document}